\documentclass{revtex4}

\usepackage{graphicx}
\usepackage{times}
\usepackage{amsmath,amssymb,enumerate}
\usepackage{bbm}
\usepackage{epsfig}

\begin{document}

\markboth{Schwarz, Bessire, Stefanov}
{VIOLATION OF A D-DIM. BELL INEQUALITY USING ENERGY-TIME ENTANGLED PHOTONS}

\title{EXPERIMENTAL VIOLATION OF A $D$-DIMENSIONAL BELL INEQUALITY USING ENERGY-TIME ENTANGLED PHOTONS}

\author{SACHA SCHWARZ$^*$, B\"ANZ BESSIRE, and ANDR\'E STEFANOV}

\address{
Institute of Applied Physics, University of Bern, \\Sidlerstrasse 5, 
CH-3012 Bern,
Switzerland\\
${}^*$sacha.schwarz@iap.unibe.ch}


\begin{abstract}
We experimentally study the violation of the CGLMP inequality for entangled 2-qubit and 2-qutrit states with different degrees of entanglement using numerically optimized measurement settings. The qudits are encoded and manipulated in the frequency spectrum of broadband energy-time entangled photons by taking into account a spatial light modulator. The latter allows to discretize the spectrum into bins. By controlling each frequency bin individually, the generation of maximally and non-maximally entangled qutrits is verified through quantum state tomography.
\end{abstract}

\maketitle

\section{Introduction}
\label{sec:introduction}
For many years people were puzzling over the question if quantum mechanics as
the underlying theory to describe non-classical phenomena is complete. After the Einstein, Podolsky, and Rosen (EPR) paradox\cite{einstein1935}, published in 1935, it was John Steward Bell who presented in 1964 a seminal theorem to tackle this question.\cite{bell1964} He accepted the EPR conclusion as a working principle and developed what is called a local hidden variable model (LHVM) in which measurement outcomes are completely predetermined and only affected locally. The assumptions of this model impose linear constraints on experimental input-output correlations in a bipartite system which have become known as Bell inequalities. An important feature in the context of an experimental test of Bell's inequality is entanglement\cite{horodecki2009}, a quantum mechanical phenomenon which occurs if the state of a particle pair cannot be written as independent single particle states. Furthermore, due to entanglement, Bell showed that it is impossible to reproduce quantum correlations in the framework of a LHVM which implies a non-local behavior of nature. But since he assumes in his work perfect correlations, the practical realization of a corresponding experiment is challenging. For this reason Clauser, Horn, Shimony, and Holt (CHSH) published in 1969 a Bell type inequality\cite{clauser1969} which allows to experimentally test a LHVM. Only a few years later, in 1972, the first experiment\cite{freedman1972} in the framework of non-locality tests was performed using entangled photons. Due to the strong resistance against decoherence photon-based systems are ideal for transmitting quantum states over large distances and provide entanglement in different degrees of freedom like their polarization\cite{kwiat1999}, transverse or orbital angular momentum (see Refs. \cite{pires2009_tomo,mair2001_tomo,langford2004,dada2011_tomo,agnew2011_tomo,fickler2012_tomo,giovannini2012_tomo}), or energy-time.\cite{law2000,thew2004,richart2012,bernhard2013} Towards the end of the century many experiments are engaged in testing the CHSH inequality.\cite{zeilinger1999} In 2002, the CHSH inquality was generalized to what is known today as the Collins-Gisin-Linden-Masser-Popescu (CGLMP) inequality\cite{collins2002} allowing for the investigation of quantum states with dimension $d \geq 2$. 

We present in this work the manifestation of quantum entanglement in bipartite 2-level as well as 3-level systems, also referred to as qubits and qutrits, by using photons entangled in the energy-time domain. Experimentally, we modulate in full coherent control\cite{peer2005} the photon spectra via a 1-dimensional spatial light modulator (SLM) allowing us to measure a CGLMP Bell parameter above the local variable limit for maximally and partially entangled 2-qubit and 2-qutrit states.\cite{bernhard2013} The corresponding measurement settings are optimized using the most general SU(3) unitary transformation.

\section{CGLMP Inequality for Bipartite Qubit and Qutrit States}
\label{sec:CGLMP}
A CGLMP experiment involves two parties renowned as Alice ($A$) and Bob ($B$). Each party performs two different measurements $A_1,A_2$ and $B_1,B_2$ on a shared 2-qudit state yielding measurement outcomes $m$ for Alice and $n$ for Bob. The total number of possible outcomes thereby depends on the dimension of the distributed quantum state, i.e.~either $m,n \in \{0, 1\}$ for qubits or $m,n \in \{0, 1, 2\}$ for qutrits. In this way, a set of joint probabilities $P(A_a = m, B_b = n)$ can be measured which determines the corresponding Bell parameter $I_{2,3}$. If     
\begin{eqnarray}
I_{2,3} & = &\vert P(A_1 = B_1) + P(B_1 = A_2 + 1) + P(A_2 = B_2) \nonumber \\ 
		& \quad & + P(B_2 = A_1) - P(A_1 = B_1 - 1) - P(B_1 = A_2) \nonumber \\ 
		& \quad & - P(A_2 = B_2 - 1) - P(B_2 = A_1 - 1 )\vert \leq 2,
\label{eq:bellIneq}
\end{eqnarray}
the correlations hidden in the set of joint probabilities can be explained by a LHVM. Whereas for a maximally entangled 2-qubit state the inequality in Eq.~\eqref{eq:bellIneq} is maximally violated by the Cirelson bound\cite{cirelson1980}, i.e.~$I_2^{max.}=2\sqrt{2}$, the Bell parameter $I_3$ is maximized to $I_3^{max.}\simeq 2.9149$ by a \textit{non}-maximally entangled 2-qutrit state.\cite{acin2002} It is therefore of interest to measure the Bell parameters $I_2$ and $I_3$ in dependence of a variable degree of entanglement. In the following we thus focus on a 2-qubit state
\begin{equation}
|\psi(\gamma)\rangle^{(2)} = \frac{1}{\sqrt{1+\gamma^2}}\left(|0\rangle_A |0\rangle_B + \gamma |1\rangle_A |1\rangle_B\right)
\label{eq:qubit}
\end{equation}
and a 2-qutrit state
\begin{equation}
|\psi(\gamma)\rangle^{(3)} = \frac{1}{\sqrt{2+\gamma^2}}\left(|0\rangle_A |0\rangle_B + \gamma |1\rangle_A |1\rangle_B + |2\rangle_A |2\rangle_B\right)
\label{eq:qutrit}
\end{equation}
with a parameter $\gamma \in [0, 1]$ quantifying the degree of entanglement. The joint probabilities in Eq.~\eqref{eq:bellIneq} are obtained via projective measurements
\begin{equation}
P_{\gamma}(A_a=m,B_b=n)\propto\left\vert\langle\chi_{m,n}\vert\psi(\gamma)\rangle^{(d)}\right\vert^2 
\label{eq:prob}
\end{equation}
with
\begin{eqnarray}\label{eq:chi_bell}
\vert\chi_{m,n}\rangle &=&\vert m\rangle_A^{a}\vert n\rangle_B^{b}\nonumber \\ &=& \left( \sum_{j = 0}^{d-1} u_j^{a}(\{\alpha_k^a\})|j\rangle_A \right) \left( \sum_{j' = 0}^{d-1} u_{j'}^{b}(\{\beta_k^b\})|j'\rangle_B \right)
\end{eqnarray}
and complex coefficients $u_j^{a,b}$. In order to maximize the Bell parameter for each value of $\gamma$, optimal measurement settings $\alpha_k^a$ and $\beta_k^b$, with $k\in \{0, 1, ..., d^2-1\}$, were determined by numerical optimization, i.e.
\begin{equation}
I_d^{\max.}(\gamma) = \max_{\{\alpha^a_k\},\{\beta^b_k\}} I_d(\{\alpha^a_k\},\{\beta^b_k\};\gamma).
\end{equation}
In Ref.~\cite{acin2002}, 
a numerical search and an analytic calculation revealed a maximal violation $I^{max.}_3$ of the CGLMP inequality under the measurement settings first given in Ref.~\cite{collins2002} and a non-maximally entangled 2-qutrit state with $\gamma=\gamma^{max.}\simeq 0.7923$. The eigenvectors of the corresponding measurement operators refer to a unitary phase operator followed by a symmetric multiport beamsplitter for each party. We, however, numerically optimize the measurement settings for each $\gamma$ incorporating the most general SU(3) unitary transformation as a measurement bases. For $d=2$ the optimality of the settings can be justified by comparing the value of the optimized Bell parameter with the corresponding outcome of the Horodecki theorem.\cite{horodecki1995} The latter provides the maximal value of the CHSH Bell parameter for a given biparite qubit state with respect to its Bell operator. In the case of $d=3$ no such theorem exists.

\section{Experimental Setup} 
\begin{figure}[t]
\centerline{\epsfig{file=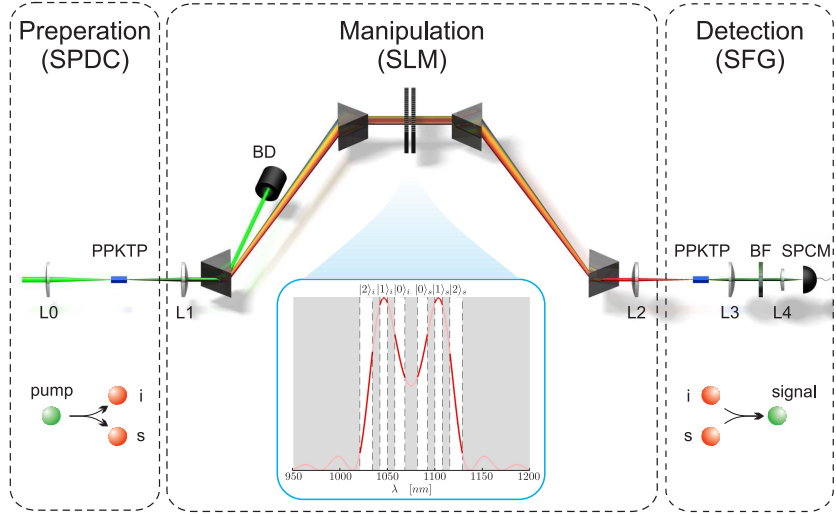,width=11cm}}
\vspace*{8pt}
\caption{Schematic view of the experimental setup. {\it Preparation}: Entangled two-photon signal is prepared via SPDC by focusing the pump beam into a nonlinear PPKTP crystal by means of lens L0 ($f = 150$ mm). {\it Manipulation}: Residue of the pump is damped by a beam dump (BD); SLM is placed in a symmetric imaging arrangement (L1 and L2 with $f=100$  mm) together with a four-prism compressor. The inset shows a simulated SPDC spectrum discretized into frequency bins introduced in Eq.~\eqref{eq:freqBins} (parts in gray refer to pixels of zero transmission).  {\it Detection}: Coincidence signal is generated via SFG and measured with a single photon counting module (SPCM) by using lenses L3 ($f = 60$ mm) and L4 ($f = 11$ mm) together with a bandpass filter (BF). The latter is used to filter out the remaining IR photons.}
\label{fig:lxil}
\end{figure}

The experimental setup can be subdivided into three parts and is schematically depicted in Fig.~\ref{fig:lxil}. By pumping a periodically poled $\text{KTiOPO}_4$ (PPKTP) crystal with a quasi-monochromatic $\text{ND:YVO}_4$ (Coherent Verdi V5) laser centered at 532 nm, collinearly phase-matched type-0 spontaneous parametric down-conversion (SPDC) is exploited. In this way, energy-time entangled photons centered at 1064 nm are prepared with a power of 1.5 $\mu$W and a spectral width of approximately 105 nm. The corresponding biphoton state can be described by first order perturbation theory 
\begin{equation}
|\psi\rangle = \int\limits_{-\infty}^{\infty}\int\limits_{-\infty}^{\infty} d\omega_i d\omega_s \Gamma(\omega_i,\omega_s)\hat{a}_i^{\dag}(\omega_i)\hat{a}_s^{\dag}(\omega_s)|0\rangle_i|0\rangle_s,
\label{eq:twoPhotonState}
\end{equation}
where we omit the leading order vacuum state. In general $\Gamma(\omega_i,\omega_s)$ is a non-separable function, denoted as the joint spectral amplitude, and $\hat{a}_{i,s}^{\dag}(\omega_{i,s})$ are operators which create the idler ($i$) and signal ($s$) photon with relative frequency $\omega_{i,s}$ by acting on a combined vacuum state. Note, that in Ref.~\cite{brida2010} the amount of entanglement of a similar state with a pulsed pump field was quantified in the frequency domain by means of an experimentally determined Fedorov parameter.%

In the subsequent manipulation part, group-velocity dispersion control is established by means of a four-prism compressor, composed of N-SF11 equilateral prisms, which at the same time serves to spatially align the spectrum of the entangled pairs at the symmetry axis of the experimental setup. A symmetric 2-lens arrangement is used to image the entangled pairs from the center of the SPDC crystal to the center of a sum-frequency generation (SFG) detection crystal. The arrangement is such that a magnification factor of 6 is achieved at the SLM plane. The SLM (Jenoptik, SLM-S640d) is used to discretize the continuous two-photon state of Eq.~\eqref{eq:twoPhotonState} into a discrete $d^2$-dimensional subspace spanned by the orthonormal product states $\vert j \rangle_i \vert k \rangle_s$ with $\vert j \rangle_{i,s} = \int_{-\infty}^{\infty} d\omega f_j^{i,s}(\omega) \, \hat{a}_{i,s}^{\dag}(\omega) \vert 0 \rangle_{i,s}$.\cite{bessire2014} Thereby, we generally obtain a 2-qudit state
\begin{equation}\label{eq:qudit_gen}
\vert \psi \rangle^{(d)} = \sum^{d-1}_{j=0} \sum^{d-1}_{k=0}\,c_{j k}\,\vert j\rangle_i \vert k \rangle_s, 
\end{equation}
with $c_{jk} = \int_{-\infty}^{\infty} d\omega_i d\omega_s f_j^{i*}(\omega_i) f_k^{s*}(\omega_s) \Gamma(\omega_i, \omega_s)$. Experimentally, a discretization scheme is applied on the SLM via a complex transfer function 
\begin{equation}\label{eq:mslm}
M^{i,s}(\omega) = \sum_{j=0}^{d-1}\,u_j^{i,s}\,f_j^{i,s}(\omega) = \sum_{j=0}^{d-1}\,\vert u_j^{i,s}\vert\,e^{i\,\phi_j^{i,s}}\,f_j^{i,s}(\omega),
\end{equation}
where each spectral component of the idler and signal photon can be independently manipulated in amplitude $\vert u_j^{i,s}\vert$ and phase $\phi_j^{i,s}$. To obtain a frequency-bin discretization the basis function $f_j^{i,s}(\omega)$ is chosen according to 
\begin{equation}
f_j^{i,s}(\omega) = \left\{
\begin{array}{cl} 
1/\sqrt{\Delta\omega_j} & \quad\text{for} \quad \vert \omega-\omega_j\vert < \Delta\omega_j/2, \\
0                     & \quad\text{otherwise}.
\end{array}
\right.
\label{eq:freqBins}
\end{equation}
If we impose for all $j,k$ that $\vert \omega_j - \omega_k \vert > (\Delta\omega_j + \Delta\omega_k)/2$, i.e.~adjacent bins do not overlap, and further take into account the continuous wave pump field in the form of a Dirac delta function, Eq.~\eqref{eq:qudit_gen} is restricted to its diagonal form
\begin{equation}\label{eq:psi_diag}
\vert \psi \rangle^{(d)} = \sum^{d-1}_{j=0}\,c_{j}\,\vert j\rangle_i \vert j \rangle_s. 
\end{equation}

Coincidences are detected through SFG which provides a time window with femtosecond temporal resolution.\cite{dayan2005} The up-converted SFG photons are finally detected by a single photon counting module (SPCM, ID Quantique, id100-50-uln). The measured signal is then described by
\begin{equation}
S = \bigg\vert\int\limits_{-\infty}^{\infty}\int\limits_{-\infty}^{\infty} d\omega_i d\omega_s \Gamma(\omega_i, \omega_s) M^{i}(\omega_i)M^{s}(\omega_s)\bigg\vert^2
\end{equation}
and is equal to the signal of a projective measurement 
\begin{equation}\label{eq:S_proj}
S = \big\vert\langle \chi | \psi \rangle^{(d)} \big\vert^2
\end{equation} 
for a direct product state
\begin{equation}\label{eq:chi}
\vert\chi\rangle=\left(\sum_{j=0}^{d-1}u^{i}_j\vert j\rangle_i\right)\left(\sum_{j'=0}^{d-1}u^{s}_{j'}\vert j'\rangle_s\right)
\end{equation}
and $\vert\psi\rangle^{(d)}$ of Eq.~\eqref{eq:psi_diag}.

By identifying the idler and signal photon with Alice ($i \leftrightarrow A$) and Bob ($s \leftrightarrow B$), respectively, the probabilities in Eq.~\eqref{eq:bellIneq} are measured by projective measurements according to Eq.~\eqref{eq:S_proj} with $\vert \chi\rangle\mapsto\vert \chi_{m,n}\rangle$ of Eq.~\eqref{eq:chi_bell}. However, since a SFG process demands a local interaction between the involved photons, the following results cannot be considered as a test of the non-local structure of nature. Though, it was shown that the Bell parameter related to $I_d$ can be used as an entanglement witness.\cite{terhal2014} As a consequence, the violation of Eq.~\eqref{eq:bellIneq} signifies entanglement in the 2-qudit state under consideration.

\section{Experimental Results}
Due to the shape of the SPDC spectrum (Fig.~\ref{fig:lxil}) the state of Eq.~\eqref{eq:psi_diag} entails unequally distributed probability amplitudes and therefore belongs to the class of non-maximally entangled qudit states. In order to experimentally obtain states according to  Eq.~\eqref{eq:qubit} and Eq.~\eqref{eq:qutrit} we first maximize the entanglement by means of the Procrustean method of entanglement concentration.\cite{bennett1996} To equate the amplitudes to $c_j=1/\sqrt{d}$ we discard photon combinations with high probabilities by adjusting the amplitudes $\vert u_j^{i,s}\vert$ of the corresponding frequency bins in the transfer function of Eq.~\eqref{eq:mslm}. After having prepared a maximally entangled state, the reduced entanglement ($\gamma<1$) is then obtained by decreasing the amplitudes $\vert u_1^{i,s}\vert$ of the bins associated with $\vert 1\rangle_i \vert 1 \rangle_s$. The states are first characterized in terms of their density matrices by performing quantum state tomography for various values of $\gamma$. Further, by applying the optimized measurement settings mentioned in Section \ref{sec:CGLMP}, the Bell parameters $I_{2,3}(\gamma)$ are measured at different degrees of entanglement.

\subsection{Quantum State Tomography}
\begin{figure}[t]
\centerline{\epsfig{file=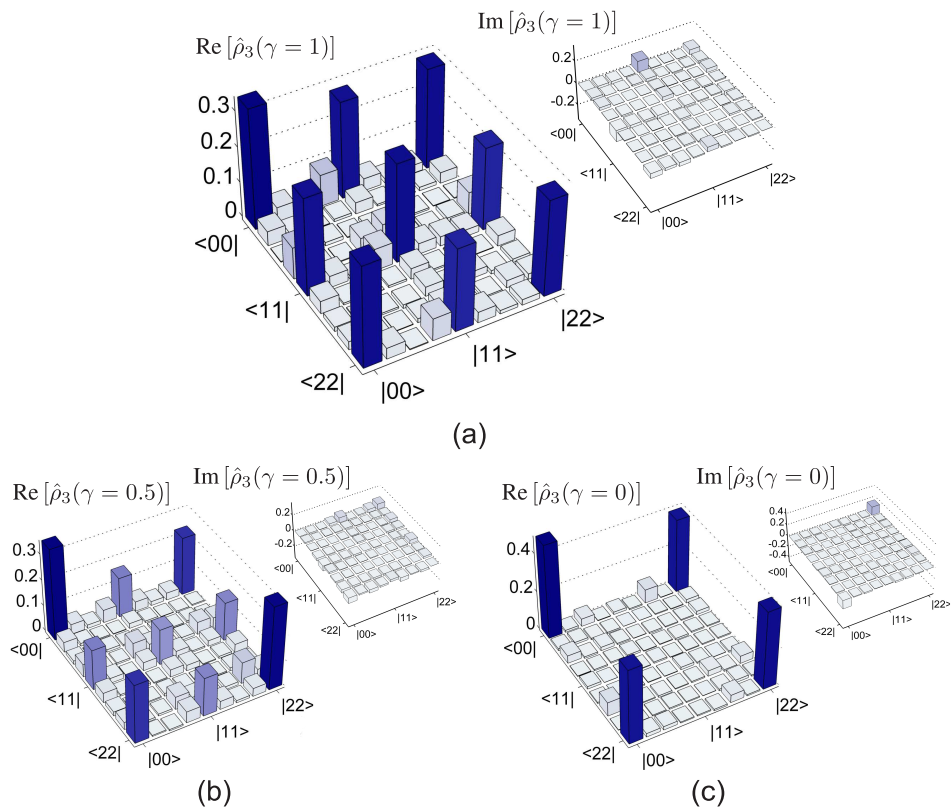,width=12.5cm}}
\vspace*{8pt}
\caption{Density matrices $\hat{\rho}_3(\gamma) = \vert\psi(\gamma)\rangle^{(3)}\,^{(3)}\langle\psi(\gamma)\vert$ for various degrees of entanglement $\gamma$ in a 2-qutrit state: (a) Maximally entangled qutrit ($\gamma=1$); (b) Non-maximally entangled qutrit (exemplary: $\gamma=0.5$); (c) $\gamma=0$ i.e.~the non-maximally entangled qutrit is reduced to a maximally entangled qubit. The corresponding fidelities are: (a) $F_3=0.83\pm 0.01$, (b) $F_3=0.65\pm 0.01$, and (c) $F_3=0.83\pm 0.01$}
\label{fig:QSTResults}
\end{figure}
Quantum state tomography (QST) fully reconstructs a quantum state in terms of its density matrix. To demonstrate the ability of our experimental setup to perform QST we focus on various 2-qutrit states according to Eq.~\eqref{eq:qutrit}. The corresponding density matrix can be expressed as a linear combination of generalized Gell-Mann matrices        
\begin{equation}\label{eq:tomodensitymat}
\hat{\rho}_3(\gamma) = \frac{1}{9}\sum_{k,l = 0}^{8} r_{kl}(\gamma) \hat{\lambda}_{k}\otimes\hat{\lambda}_{l},\ \ r_{kl}\in \mathbb{R},
\end{equation}
and the $d$-dimensional identity matrix $\hat{\lambda}_{0}$.\cite{thew2002}  
To determine the coefficients $r_{kl}(\gamma)$ we chose $\vert \chi\rangle$ in Eq.~\eqref{eq:chi} according to a tomographically complete set of projective measurements provided in Ref.~\cite{agnew2011_tomo}. In order to retrieve a positive semidefinite $\hat{\rho}_3(\gamma)$ out of the tomographic measurements, we employ a maximum likelihood estimation method for high dimensional states established in the aforementioned Reference. The resulting 2-qutrit density matrices for $\gamma\in [0,0.5,1]$ are depicted in Fig.~\ref{fig:QSTResults}. The fidelity\cite{josza1994} $F_d=[\mbox{Tr}(\sqrt{\sqrt{\hat{\rho}_{d,T}}\hat{\rho}_{d}\sqrt{\hat{\rho}_{d,T}}})]^2$ is calculated with respect to the target density matrix $\hat{\rho}_{d,T}$ given by the pure state in Eq.~\eqref{eq:qutrit}. The 2$\sigma$ uncertainties on $F_d$ are calculated through a
Monte Carlo method by randomly adding normally distributed noise to each measurement outcome and recomputing the fidelity. 

\subsection{Bell Parameter Measurement}
\begin{figure}[t]
\centerline{\epsfig{file=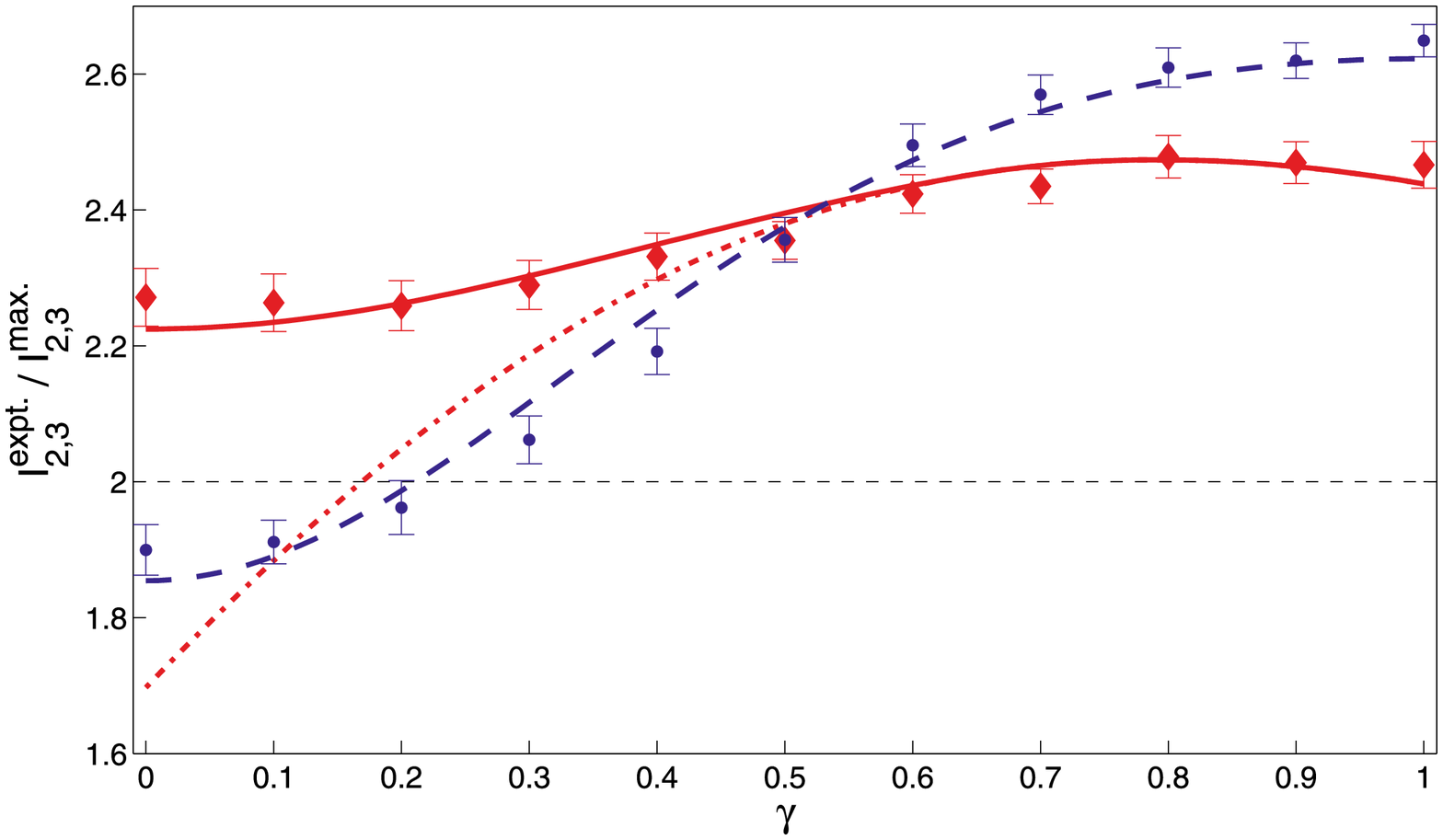,width=12.5cm}} 
\vspace*{8pt}
\caption{Bell parameter $I_d$ in dependence of the entanglement parameter $\gamma$. The experimental Bell parameter $I^{\text{expt.}}_2$ is depicted with blue dots and $I^{\text{expt.}}_3$ with red diamonds. The $1\sigma$ uncertainties are calculated assuming Poisson statistics on background-subtracted coincidence counts. The theoretically predicted Bell parameters $I_2^{max.}(\gamma)$ (dashed blue line) and $I_3^{max.}(\gamma)$ (solid red line) are scaled with their corresponding mixing parameter. The additional red dot-dashed line represents the theoretical Bell parameter $I_3^{max.}(\gamma)$ incorporating symmetric three-port beamsplitters. We experimentally determine the mixing parameters to be $\lambda^{\text{expt.}}_2=0.927\pm0.045$ and $ \lambda^{\text{expt.}}_3=0.849\pm0.041$ with  1$\sigma
$ uncertainties. The (horizontal) dashed black line indicates the local variable limit.}
\label{fig:results}
\end{figure}
In order to experimentally determine $I_{2,3}^{max.}(\gamma)$ of Eq.~\eqref{eq:bellIneq} for a specific value of $\gamma$, a number of 16 and 36 different projective measurements are required  for qubit- and qutrit-based systems, respectively. For various values of $\gamma$, Fig.~\ref{fig:results} presents experimental data, i.e.~$I_{2,3}^{expt.}(\gamma)$, together with theoretical predictions for the Bell parameters $I_2^{max.}(\gamma)$ and $I_3^{max.}(\gamma)$ using optimized measurements settings. The theoretical curves are thereby scaled to the experimental data using a symmetric noise model 
\begin{equation}
\hat{\rho}_d^{\text{sn}}(\gamma) =  \lambda_d \vert \psi(\gamma) \rangle^{(d)} {}^{(d)}\langle \psi(\gamma) \vert + \frac{1-\lambda_d}{d^2} \mathbf{1}_{d^2},
\end{equation} 
where the mixing parameter $\lambda_d$ and the $d^2$-dimensional identity operator $\mathbf{1}_{d^2}$ are used. Therefore, deviations from a pure state due to white noise and experimental imperfections can be quantified by $\lambda_d \in \left[0, 1\right]$. The value of the Bell parameter for $\hat{\rho}_d^{\text{sn}}(\gamma)$ then scales as $\lambda_d I_{d}^{max.}(\gamma)$. 

We observe a smaller sensitivity to $\gamma$ of the Bell parameter for qubits compared to qutrits which is in accordance with theoretical predictions. As discussd, in both systems, higher Bell parameter values are obtained after optimizing the corresponding experimental settings. To motivate the use of the optimized SU(3) operators, we additionally depict in Fig.~\ref{fig:results} the Bell parameter $I_3^{max.}(\gamma)$ now optimized with respect to a symmetric multiport beamsplitter combined with a unitary phase operator as discussed in Ref.~\cite{acin2002} (red dot-dashed line). In this case the measurement settings $\alpha_k^{a}$ and $\beta_{k}^{b}$ to be optimized solely consist of phase angles whereas the full SU(3) involves also optimized amplitudes in the projection states.
The red dot-dashed line shows the necessity of the SU(3) optimization procedure since smaller values of the Bell parameter are obtained, especially for small $\gamma$, compared to the SU(3) case (solid red line).

\section{Conclusions \& Outlook}
In summary, we have demonstrated the quantum state reconstruction of maximally and non-maximally entangled qutrits encoded in frequency bins by a SLM. The latter provides a versatile tool to manipulate a quantum state which allowed to study the CGLMP Bell inequality for a 2-qubit and a 2-qutrit state with a variable degree of entanglement. The Bell test measurements presented here can be readily extended to a more general class of qutrits and projection measurements, and, thus, can help to tackle the experiments imposed by the numerical Bell test studies in Ref.~\cite{gruca2012}. Apart from the CGLMP inequality, there exists a more general class of Bell inequalities which demands $d$-input settings per party instead of only two.\cite{liang2009} The experimental examination of such a multisetting Bell inequality for qutrits is currently under study. 

\section*{Acknowledgments}

This research was supported by the grant PP00P2\_133596 funded by the Swiss National Science Foundation.

\end{document}